\documentclass[11pt]{article}
\usepackage{moriond,epsfig}

\def\Journal#1#2#3#4{{#1} {\bf #2}, #3 (#4)}

\def\NIMA{{\em Nucl. Instrum. Methods} A}
\def\NPA{{\em Nucl. Phys.} A}

\def\PRL{\em Phys. Rev. Lett.}
\def\PRC{{\em Phys. Rev.} C}

\def\be{\begin{equation}}
\def\ee{\end{equation}}
\def\bea{\begin{eqnarray}}
\def\eea{\end{eqnarray}}

\def\pt{$p_{T}$}
\begin{document}
\vspace*{4cm}
\title{Direct Photons and Photon-Hadron Correlations at PHENIX}

\author{B.~Sahlmueller\footnote{for the PHENIX collaboration} }

\address{Department of Physics and Astronomy, Stony Brook University,\\
Stony Brook, NY 11790, USA}

\maketitle\abstracts{
  Direct photons are a powerful tool to study the hot and dense matter created in heavy-ion collisions at RHIC, since they are created in the different stages of the collision. Since they do not interact via the strong force, they can travel through the hot and dense matter mostly unaffected. The PHENIX experiment has measured direct photons using different methods, over a broad range of transverse momentum (\pt), in different collision systems, and at different energies. These measurements help interpreting the measurement of hadrons as well as understanding the temperature of the created quark-gluon plasma (QGP). The azimuthal anisotropy of direct photons may shed light on the thermalization time of the medium. Using direct photons to tag jets is a crucial tool to understand the energy loss of scattered partons in the medium.
}


The experimental program at the Relativistic Heavy-Ion Collider has found evidence for the creation of a quark-gluon plasma in collisions of Au nuclei at center-of-mass energies of 200~GeV per nucleon.~\cite{Adcox:2004mh} One of the crucial signatures are direct photons that are produced in such collisions and can traverse the created QGP mostly unaffected.

Direct photons are defined as photons that are not from decays of hadrons, such as $\pi^0$ or $\eta$. These photons are produced in different stages of a heavy-ion collision over a broad range of transverse momentum. At large and intermediate transverse momentum (\pt) they are produced mainly from initial hard scattering processes of the colliding quarks or gluons such as $q+g \rightarrow q+\gamma$ or $q+\bar{q} \rightarrow g+\gamma$, as bremsstrahlung emitted by a scattered parton, from the fragmentation of such quarks and gluons, or from the interaction of a scattered parton with the strongly interacting medium created in such collisions~\cite{Gale:2009gc}. In the hard scattering processes, a parton is emitted opposite to the photon, that will subsequently fragment into a hadronic jet. Hence, the energy of the jet is balanced with the energy of the direct photon on the opposite side. At low \pt, the medium can emit thermal direct photons directly. Their \pt~ distribution depends on the average temperature of the medium.~\cite{Stankus:2005eq}
To account for nuclear effects, the direct photon yield in Au+Au collisions is compared to the cross section in $p+p$ collisions with the help of the nuclear modification factor which is defined as
\begin{eqnarray}
  R_{AA} = \frac{d^2N/dp_Tdy|_{AuAu}}{\left<T_{AA}\right>d^2\sigma^{pp}/dp_Tdy}\mbox{ ,}
\end{eqnarray}
where $\left<T_{AA}\right>$ is the nuclear overlap function.

The azimuthal anisotropy of direct photons is sensitive to the different production processes, it is measured in terms of the anisotropy parameter $v_2$ which is the second harmonic of the Fourier expansion of the azimuthal distribution. The elliptic flow of thermal direct photons is sensitive to the thermalization time $\tau_0$ of the QGP, small $\tau_0$ would lead to small $v_2$.~\cite{Chatterjee:2008tp}


Direct photons can be measured using different subsystems of the central arm of the PHENIX detector, with different methods of measurement. The PHENIX detector is described elsewhere.~\cite{Adcox:2003zm} Photons can be measured directly using the Electromagnetic Calorimeters, the decay photons from $\pi^0$ and other mesons are subtracted statistically from the measured inclusive photon sample, charged hadrons and electrons are rejected with the help of the central arm tracking detectors. This method is described in more detail in earlier publications~\cite{Adler:2005ig}. It is most feasible at high $p_T$, at low $p_T$ the signal to background ratio gets too small to make a significant measurement. Therefore, two other methods have been developed, using the electron ID capabilities of the PHENIX detector and measuring direct photons indirectly through conversions.

The so-called internal conversion method uses the idea that virtual direct photons are produced in conjunction with real direct photons, and convert into low mass $e^+e^-$ pairs, in a process similar to the $\pi^0$ Dalitz decay. The method benefits from the limited phase space of such pairs from the $\pi^0$ Dalitz decay, hence the signal to background ratio is improved compared to the direct calorimeter measurement at low \pt. The method is based on the measurement of $e^+e^-$ pairs, the background is removed using like-sign pairs, the resulting distribution is compared to a cocktail of dilepton pairs that includes all expected hadronic sources. An excess over that cocktail at low invariant mass is then interpreted as a virtual direct photon signal. A more detailed description of this method is given in~\cite{Adare:2008fqa}.

A new method has been developed to measure direct photons through external conversions in the detector material. The back plane of the PHENIX Hadron-Blind Detector (HBD), which was installed during the 2007 RHIC run for commissioning, offers a well-defined conversion point for photons, about 60~cm away from the interaction point, with a radiation length of about $4\%X_0$. To account for the wrongly reconstructed opening angle of such conversion pairs, that leads to an apparent invariant mass, an alternate track model was developed that assumes the origin of the pairs at the HBD back plane. Using this model moves the peak of the invariant mass from about 12~MeV/$c^2$ to 0, it also helps separating the conversion pair from pairs from Dalitz decays. Using this method, the misidentification rate of conversion photons is found to be less than 3\%. This method is described in more detail in~\cite{Petti:2011a}.

\begin{figure}
\centering
\includegraphics[height=4.9cm]{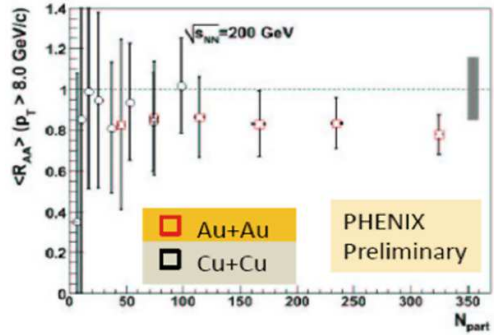}
\caption{Average nuclear modification factor $<R_{AA}>$ for direct photons in Au+Au and Cu+Cu collisions at 200~GeV, plotted versus $N_{part}$.
\label{fig:raa_int}}
\end{figure}
The nuclear modification factor $R_{AA}$ has been measured in 200 GeV Au+Au collisions and was found to be consistent with unity for most \pt.~\cite{Isobe:2006aa} This result is a confirmation that $\left<T_{AA}\right>$ scaling works, since at a first approximation, photons are not affected by the QGP. However, there appears to be a possible suppression of photons at $p_T > 15$~GeV/$c$, which is not fully understood. Such a suppression could be an initial state effect, for example an effect of the different isospin composition of the proton and of the gold nuclei, and would thus be visible in d+Au collisions at the same energy. The isospin effect would also be visible in Au+Au collisions at 62.4~GeV, at lower \pt, since it scales with $x_T = 2p_T/\sqrt{s_{NN}}$. PHENIX has done both measurements, with the 2003 and 2004 datasets, respectively, and they are both statistically too limited to draw any conclusion. Direct photons have also been measured in Cu+Cu collisions at 200~GeV, the averaged nuclear modification factor is compared with Au+Au collisions in Fig.~\ref{fig:raa_int} and found to be consistent for similar numbers of participants.

\begin{figure}
\centering
\epsfig{figure=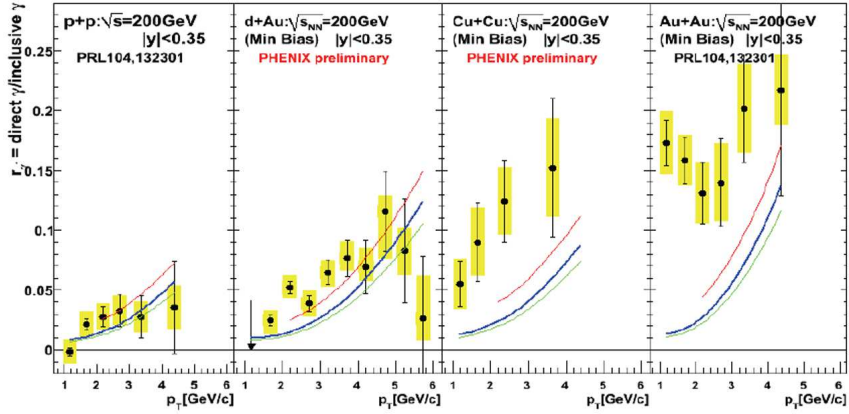,height=5.5cm}
\caption{Fraction of direct photons over inclusive photons, for different collision systems, at center-of-mass energy of 200 GeV: $p+p$, $d$+Au. Cu+Cu, Au+Au (from left to right).
\label{fig:dirgam_virt}}
\end{figure}
The virtual direct photon measurement has been done for four different collision systems at 200~GeV collision energies, a $p+p$ measurement works as a baseline to understand the other measurements, a $d$+Au measurement is used to look for effects of cold nuclear matter, and the measurements in Au+Au and Cu+Cu are used to study properties of the QGP, also with respect to different system sizes. The ratio of direct photons and inclusive photons is shown in Fig.~\ref{fig:dirgam_virt} for all four collision systems. While the $p+p$ measurement shows that pQCD agrees well with the data, there is a clear excess in the Au+Au data. The excess is smaller in Cu+Cu and does not appear in the $d$+Au measurement, which gives evidence that it is indeed a final state effect. When fitting the excess over scaled pQCD of direct photon yield in Au+Au with an exponential, the average temperature of the medium can be extracted as the inverse slope of the function. It is found to be $221 \pm 19 (stat) \pm 19 (sys)$ MeV.

The external conversion measurement is still on its way, its final goal is to measure the elliptical flow of thermal photons. The method has been found to produce reliable results for the inclusive photon $v_2$, this measurement agrees well with an earlier measurement using the EMCal.~\cite{Petti:2011a}

\begin{figure}
\begin{minipage}[hbt]{8cm}
\centering
\epsfig{figure=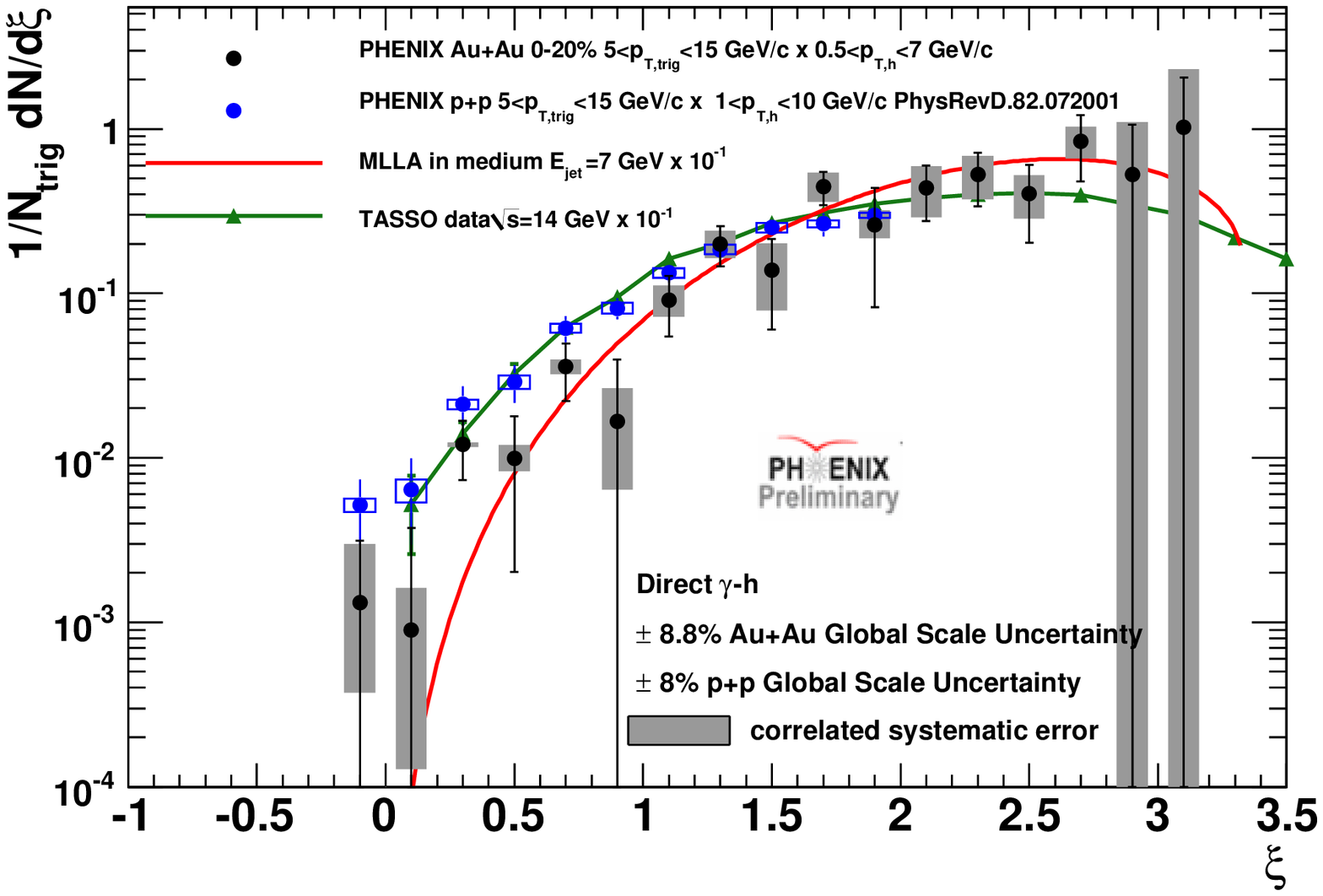,height=4cm}
\caption{$\xi$ distribution for PHENIX Au+Au (black circles) and $p+p$ (open blue circles), compared to a MLLA prediction (red line) and TASSO data (green triangles). 
\label{fig:xi_distri}}
\end{minipage}
\hspace{\fill}
\begin{minipage}[hbt]{8cm}
\centering
\epsfig{figure=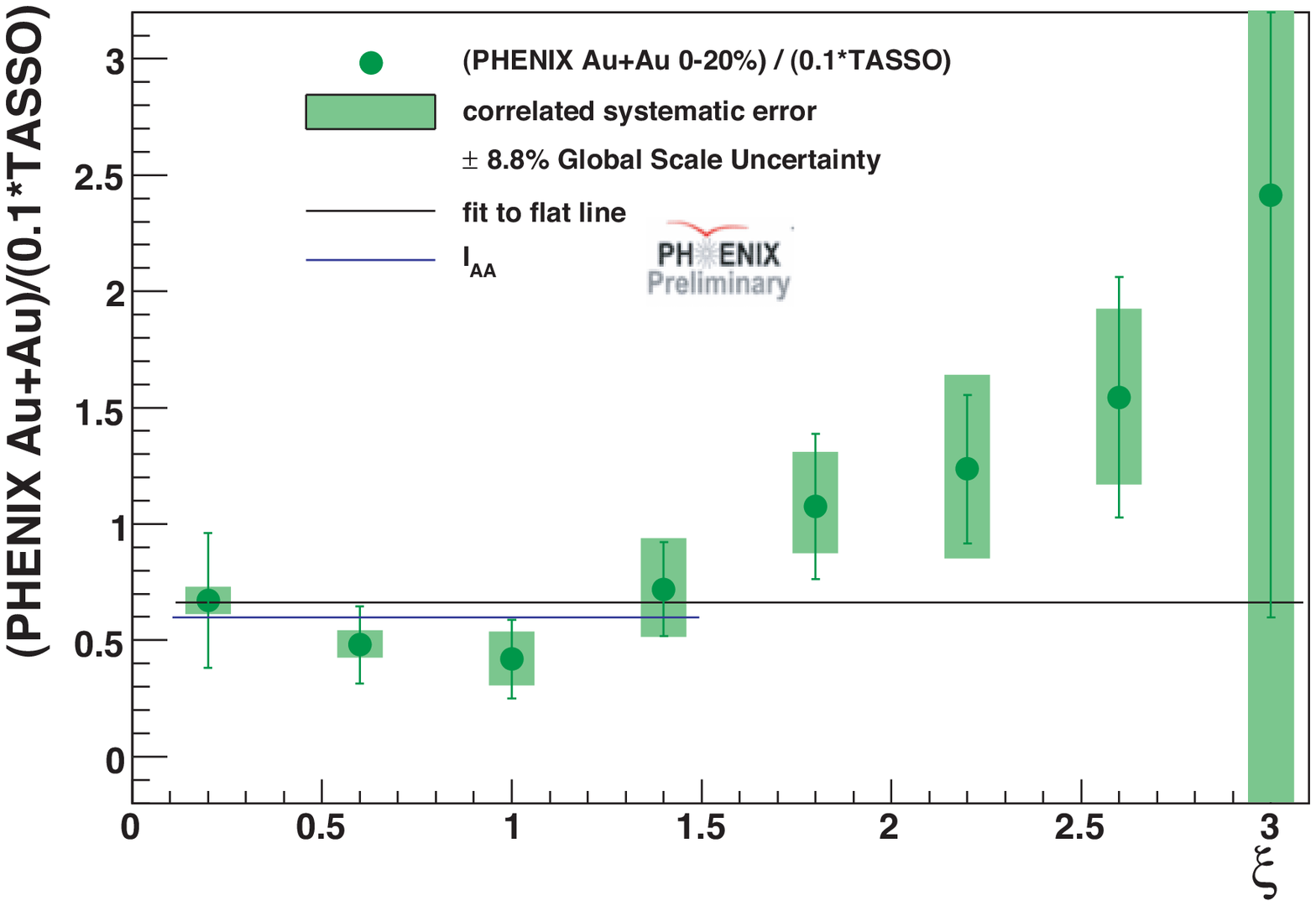,height=4cm}
\caption{Ratio of Au+Au to TASSO data scaled by a factor of 10. The black line is a fit to the ratio, the blue line is a fit to the $I_{AA}$ using PHENIX $p+p$ as reference.
\label{fig:iaa}}
\end{minipage}
\end{figure}

Another way of using direct photons to study the QGP is using direct photons as trigger to tag jets, for photons from hard scattering processes balance the energy from jets on the opposite side. To measure the modification of the jet, PHENIX uses correlations of direct photons and hadrons. Since direct photons cannot be measured on an event-by-event basis, first the yield per trigger ($Y_{inc}$) of inclusive photons and hadrons is measured as well as the yield per trigger of the $\pi^0$ and hadrons. The yield per trigger for decay photons ($Y_{dec}$) is calculated based on the $\pi^0$-hadron correlation, and finally, using the ratio of inclusive and decay photons, $R_{\gamma}$, the yield per event for direct photons is calculated as
\begin{eqnarray}
Y_{dir} = \frac{R_{\gamma}Y_{inc} - YÐ{dec}}{R_{\gamma}-1}\mbox{ .}
\end{eqnarray}
A much more detailed description of this analysis is given in another publication~\cite{Connors:2011a}.

The integrated yield on the away side is calculated for both $p+p$ and Au+Au data, it is plotted as a function of the fragmentation variable $z = p_T^h / p_T^{trigger}$ to show the fragmentation function. An alternative way of plotting is showing the distribution as a function of $\xi = -\ln(x_E)$, where $x_E = p_T^h \cos(\Delta\phi) / p_T^{trigger}$, this plot is shown in Fig.~\ref{fig:xi_distri} for both $p+p$ and Au+Au, including also $e^+e^-$ data from TASSO~\cite{Braunschweig:1990a} and a theoretical prediction from the Modified Leading Logarithmic Approximation (MLLA) in the medium~\cite{Borghini:2005a}, the latter two curves are scaled down arbitrarily by a factor of 10 to account for the limited PHENIX acceptance.\\
Dividing the yield in Au+Au by the yield in $p+p$, $I_{AA}$ can be calculated, a variable to quantify effects of the medium in Au+Au. Since the scaled TASSO agrees with the PHENIX $p+p$ data, and since it extends to higher $\xi$, it can be used as a baseline instead of the $p+p$. The $I_{AA}$ like ratio calculated with the TASSO data is shown in Fig.~\ref{fig:iaa}. The shape is consistent with a flat line and a suppression of the Au+Au yield below $\xi=1.8$, but the points above indicate a change in shape and suggest an enhancement at highest $\xi$ values, which can be interpreted as the response of the medium to the lost energy.

In summary, direct photons are a powerful tool to study the QGP created in ultrarelativistic heavy-ion collisions. The measurement of direct photons shows that binary scaling works when comparing heavy-ion collisions to baseline $p+p$ or $d$+Au collisions. Photons are also emitted from the medium directly or through interaction of partons with the medium. An excess of direct photons at low $p_T$ can be interpreted as a thermal signal from the QGP, and an average medium temperature of $221 \pm 19 (stat) \pm 19 (sys)$ MeV could be extracted. The measurement of flow of direct photons could further disentangle different photon production mechanisms. Direct photons are also crucial for probing the matter with direct photon-hadron correlations where the photon balances the jet energy. This measurement showed suppression of the away side and a shape suppression of the fragmentation function at high $\xi$ which can be related to the medium response to energy loss.


\begin{thebibliography}{99}
\bibitem{Adcox:2004mh} K.~Adcox~et~al., \Journal{\NPA}{757}{184}{2005}
\bibitem{Gale:2009gc} C.~Gale, arXiv:0904.2184 (hep-ph) (2009)
\bibitem{Stankus:2005eq} P.~Stankus, \Journal{Ann.Rev.Nucl.Part.Sci.}{55}{517}{2005}
\bibitem{Chatterjee:2008tp} R.~Chatterjee, D.~Srivastava, \Journal{\PRC}{79}{021901}{2009}
\bibitem{Adcox:2003zm} K.~Adcox~et~al., \Journal{\NIMA}{499}{469}{2003}
\bibitem{Adler:2005ig} S.~S.~Adler~et~al., \Journal{\PRL}{94}{232301}{2005}
\bibitem{Adare:2008fqa} A.~Adare~et~al., \Journal{\PRL}{104}{132301}{2010}
\bibitem{Petti:2011a} R.~Petti, conference proceedings, WWND 2011, to be published (2011)
\bibitem{Isobe:2006aa} T.~Isobe, \Journal{J.Phys.G}{34}{1015}{2007}
\bibitem{Connors:2011a} M.~Connors, \Journal{\NPA}{855}{335}{2011}
\bibitem{Braunschweig:1990a} W.~Braunschweig~et~al., \Journal{Z.Phys.C}{47}{187}{1990}
\bibitem{Borghini:2005a} N.~Borghini and A.~Wiedemann, arXiv:hep-ph-0506218 (2005)
\end{thebibliography}

\end{document}